\documentclass{article}[12pt]
\usepackage{geometry}[left=1.25in, right=1.25in, top=1.25in, bottom=1.25in]
\usepackage{setspace}
\usepackage{graphicx} 
\usepackage{amsmath, amssymb}
    \allowdisplaybreaks
\usepackage{amsthm}
    \newtheoremstyle{customstyle}
      {1.5\topsep} 
      {1.5\topsep} 
      {\itshape} 
      {} 
      {\bfseries} 
      {.} 
      {.5em} 
      {} 
    \theoremstyle{customstyle}
        \newtheorem{theorem}{Theorem}[]
        \newtheorem*{theorem*}{Theorem} 
        \newtheorem*{definition}{Definition}
        \newtheorem*{proposition}{Proposition}
        
        \newtheorem*{assumption}{Assumption}
        \newtheorem*{corollary}{Corollary}
\usepackage{tikz}
\usepackage{pgfplots}
    \pgfplotsset{compat=1.16}
\usepackage{appendix}
\usepackage{enumitem}
    \setlist[enumerate]{topsep=2pt, itemsep=2pt, parsep=0pt}
\usepackage{xcolor}
\usepackage{natbib}
	\setcitestyle{authoryear,round,comma,aysep={},yysep={,},notesep={}}
\usepackage{authblk}

\title{The matching problem with linear transfers is equivalent to a hide-and-seek game\thanks{We thank Sylvain Sorin for helpful suggestions.}}
\author[1]{A.\ Galichon\thanks{Economics Department, FAS, and Mathematics Department, Courant Institute, New York University; and Economics Department, Sciences Po. Email: ag133@nyu.edu. Galichon gratefully acknowledges funding from ERC grant CoG-866274.},
A.\ Jacquet\thanks{Economics Department, Sciences Po. Email: antoine.jacquet@sciencespo.fr. Funding from ERC grant CoG-866274 is gratefully acknowledged.}}
\date{April 2024}

\begin{document}

\vspace{-150pt}

\maketitle

\begin{abstract}
    \noindent Matching problems with linearly transferable utility (LTU) generalize the well-studied transferable utility (TU) case by relaxing the assumption that utility is transferred one-for-one within matched pairs.
    We show that LTU matching problems can be reframed as nonzero-sum games between two players, thus generalizing a result from \citeauthor{vonNeumann1953}.
    The underlying linear programming structure of TU matching problems, however, is lost when moving to LTU.
    These results draw a new bridge between non-TU matching problems and the theory of bimatrix games, with consequences notably regarding the computation of stable outcomes.
\end{abstract}



\section{Introduction}
\label{sec:introduction}

In a 1953 paper, John \citeauthor{vonNeumann1953} established an equivalence between the optimal assignment problem and a two-person, zero-sum game known as hide-and-seek.
The equivalence is meant in the following sense: one can recover the solutions to the optimal assignment problem from the Nash equilibria of the hide-and-seek game, and vice-versa.
Later on, \citet{Shapley-Shubik1971} used linear programming duality to also reframe the assignment problem as a matching problem with transferable utility (TU).
Overall, the combination of these two results created an equivalence between matching problems with TU and zero-sum hide-and-seek games. 

In this note we generalize this result to a larger class of matching problems, whose set of feasible utilities satisfy a condition that we call linearly transferable utility (LTU).
LTU essentially relaxes the TU assumption that utility must be transferred at a one-for-one rate within all matched pairs.
Instead, it allows for different matched pairs to have different utility transfer rates.
LTU holds for instance in a matching problem between workers and employers with differential taxation across worker and job types.

We show that matching problems with LTU are equivalent to two-player games which are a nonzero-sum generalization of von Neumann's hide-and-seek game.
We present this nonzero-sum hide-and-seek game, and we characterize the equivalence between this game and the matching problem with LTU.
Finally, we also show how this equivalence extends to a larger class of matching problems with LTU, whereby individuals can match freely in arrangements of up to $N$ members. (This notably includes the roommate problem or the one-to-many matching problem.) Our result has implications regarding both the existence of solutions to the matching problems with LTU and their computation.
Indeed, the existence of solutions can now be seen as a direct consequence of the existence of Nash equilibria in finite games.
Furthermore, the computation of solutions to the matching problem can now rely on the tools and machinery available for computing Nash equilibria in finite games.


\paragraph{}
This note's first contribution is to the understanding of non-TU matching problems, and specifically the link between LTU matching problems, bimatrix game theory, and mathematical programming.
Apart from the already-mentioned contribution by \citet{vonNeumann1953}, efforts date back at least to \citet{Dantzig1951}, who showed how a linear programming problem could be reframed as a two-person zero-sum game, and vice-versa.
Dantzig's proof was completed recently by \citet{Adler2013} and \citet{vonStengel2023}.
Here we show that matching problems with LTU can also be reframed as games, but \emph{not} as linear programs.

The second contribution is to the existence of solutions in matching problems.
The existence of such solutions was already demonstrated (in a more general nonlinear setting, see \citealt{Roth-Sotomayor1990}, ch.\ 9) by \citet{Crawford-Knoer1981} using a limiting argument from the discrete case, and then directly by \citet{Quinzii1984}.
\citet{Alkan-Gale1990} provide a simple proof in the piecewise-linear case.
Recently, \citet{Noldeke-Samuelson2018} also proved the existence of such solutions, this time using the link between matching problems and Galois connections.
What we show is that in the specific case of LTU matching problems, there is an alternative route to prove the existence of solutions, through the existence of Nash equilibria in finite games.

The third contribution concerns the computation of these solutions.
\citet{Kelso-Crawford1982} and \citet{Alkan1989} provided algorithms to find solutions based on their respective proofs of existence.
By reframing LTU matching problems as finite games, we open the way to leverage the existing computational methods from game theory (see for instance \citealt{vonStengel2002}) in order to solve matching problems.
This is important notably for the empirical study of matching models when utility is not perfectly transferable (see for instance \citealt*{Legros-Newman2007, Chiappori-Reny2016, Galichon-Kominers-Weber2019}).

Finally, it is worth noting that our result differs from that of \citet{Scarf1967} who, along with later works like \citet{Kelso-Crawford1982} or \citet{Hatfield-Milgrom2005}, examined the problem of finding a stable outcome in an economy with a finite number of potential contracts.
In contrast, our LTU framework (which generalizes von Neumann's TU framework) involves a continuum of contracts.
Interestingly, the method of \citet{Scarf1967} also involves two-person games -- although not hide-and-seek.


\paragraph{}
The note is structured as follows.
Section \ref{sec:LTU_matching} presents the matching problem with linearly transferable utility and the usual stability conditions.
Section \ref{sec:hide_seek} introduces the generalization of \citeauthor{vonNeumann1953}'s hide-and-seek game, specifies the equivalence between this game and the LTU matching problem, and discusses the implications.
Section \ref{sec:exchangeability_TU} shows how the relationship between matching problems with TU and linear programming unravels in the LTU case, using the notion of solution exchangeability.

\section{Matching with linearly transferable utility}
\label{sec:LTU_matching}

We consider the problem of matching workers with jobs as an illustration.
Workers belong to a finite number of types $x \in X$, and jobs to a finite number of types $y \in Y$.
The mass of workers of type $x$ is $n_x$, and the mass of jobs of type $y$ is $m_y$.
We do not impose that $\sum_x n_x$ be equal to $\sum_y m_y$, so that there can be an excess of workers or an excess of jobs on the market.

In the well-studied transferable utility (TU) case, it is assumed that the match between a worker $x$ and a job $y$ creates a surplus $\Phi_{xy}$.
The worker and the employer then agree on a division of this surplus according to the feasibility constraint
\begin{equation}
    u_x + v_y = \Phi_{xy},
    \label{eq:TU}
\end{equation}
where $u_x$ is the utility of the worker and $v_y$ that of the employer.
According to this constraint, any increase in the utility of the worker must come at the expense of the utility of the employer, and vice-versa.
Furthermore, utility is always transferred at the same one-for-one rate between worker and employer, no matter the match $xy$.

Here we introduce a generalization of the TU case, which we call \emph{linearly transferable utility} (LTU).
In the LTU case, utility is still transferred at a constant rate between worker and employer, but this rate now differs across matches.

\begin{definition}
The matching problem satisfies the linearly transferable utility (LTU) condition if for all $x \in X$ and $y \in Y$, there exist numbers $\lambda_{xy} \in (0,1)$ and $\Phi_{xy}$ such that the feasibility constraint of the match $xy$ can be written
\begin{equation}
    \lambda_{xy} \, u_x + (1-\lambda_{xy}) \, v_y = \Phi_{xy}/2.
    \label{eq:LTU}
\end{equation}
\end{definition}

\paragraph{}
Thus in the LTU case, utility can be transferred at a match-specific rate of $\lambda_{xy}$-for-$(1-\lambda_{xy})$;
that is, for the worker's utility to increase by $\lambda_{xy}$, the employer's utility must decrease by $(1-\lambda_{xy})$.
Clearly, the TU case obtains when $\lambda_{xy} = 1/2$ for all $xy$.
Contrary to the TU case however, the joint output of the match $\Phi_{xy}/2$ cannot be interpreted as surplus anymore in the LTU case.
This is because the value of the output itself, both for the worker and for the employer, depends on whom they are paired with.
Indeed, one unit of the output is worth $1/\lambda_{xy}$ to the worker, and $1/(1-\lambda_{xy})$ to the employer.

\paragraph{\textnormal{\itshape Example.}}
Suppose that hiring worker $x$ produces a value $S_{xy}$ for employer $y$.
The employer can compensate the worker by paying them a gross wage $w$.
This gross wage is taxed at the rate $\tau_{xy}$, which depends on the worker type $x$ (e.g.\ their age, marital status, disability), and on the job type $y$ (e.g.\ the job's industry).
After tax, the worker receives the net wage $(1-\tau_{xy}) w$.
The utilities are therefore $u_x = (1-\tau_{xy}) w$ and $v_y = S_{xy} - w$, and the feasibility constraint of the match is $u_x /(1-\tau_{xy}) + v_y = S_{xy}$. Multiplying this constraint by $(1-\tau_{xy})/(2-\tau_{xy})$, we see that it is a LTU problem with $\lambda_{xy} = 1/(2-\tau_{xy})$ and $\Phi_{xy}/2 = (1-\tau_{xy}) S_{xy}/(2-\tau_{xy})$.

\paragraph{}
We now provide the characterization of a stable outcome under the linearly transferable utility condition.
An outcome $(\mu, u, v)$ is a vector which specifies, for any $x$ and $y$, the mass $\mu_{xy}$ of workers $x$ matched to jobs $y$, as well as the utility $u_x$ obtained by workers $x$ and the utility $v_y$ obtained by employers $y$.
We set the utility of being unmatched to $0$ for all workers and employers, so that we can focus attention on outcomes such that $\mu_{xy}, u_x, v_y \geq 0$ for all $x$ and $y$.

LTU is a particular case of imperfectly transferable utility (ITU)
(see \citealt{Chiappori2017}, ch.\ 7; also called a ``generalization of the assignment model'' in \citealt{Roth-Sotomayor1990}).
Hence, an outcome $(\mu, u, v)$ is said to be stable if and only if the following conditions (\ref{eq:LTU_stability1})–(\ref{eq:LTU_stability6}) are satisfied:
\begin{align}
    \lambda_{xy} \, u_x + (1-\lambda_{xy}) \, v_y &\geq \Phi_{xy}/2
    &\qquad \forall x \in X, y \in Y \label{eq:LTU_stability1} \\
    \textstyle \sum_y \mu_{xy} &\leq n_x
    &\qquad \forall x \in X \label{eq:LTU_stability2} \\
    \textstyle \sum_x \mu_{xy} &\leq m_y
    &\qquad \forall y \in Y \label{eq:LTU_stability3}
\end{align}
\vspace{-25pt}
\begin{align}
    \mu_{xy} > 0 &\implies \lambda_{xy} \, u_x + (1-\lambda_{xy}) \, v_y = \Phi_{xy}/2
    \label{eq:LTU_stability4} \\
    u_x > 0 &\implies \textstyle \sum_y \mu_{xy} = n_x 
    \label{eq:LTU_stability5} \\
    v_y > 0 &\implies \textstyle \sum_x \mu_{xy} = m_y.
    \label{eq:LTU_stability6}
\end{align}

Condition (\ref{eq:LTU_stability1}) guarantees that $xy$ is not a \emph{blocking pair}, that is, a worker $x$ and an employer $y$ cannot both improve on their current level of utility by pairing together.
Conditions (\ref{eq:LTU_stability2})--(\ref{eq:LTU_stability3}) are feasibility constraints on the population.
Condition (\ref{eq:LTU_stability4}) means that the feasibility constraint (\ref{eq:LTU}) of the match $xy$ must bind whenever pairs $xy$ are formed.
Finally, conditions (\ref{eq:LTU_stability5})--(\ref{eq:LTU_stability6}) mean that if a type of worker or employer has strictly positive utility, then there can be no individual of that type who remains single.

In the next section we will assume that match outputs are strictly positive.
\begin{assumption}[Positive outputs]
$\Phi_{xy} > 0$ for all $x \in X$ and $y \in Y$.
\end{assumption}
While not necessary for our result, positive outputs keep the exposition more concise.
Along with a generalization of the model to the many-to-one case, appendix \ref{apx:many-to-one} shows how to account for nonpositive outputs.

\section{The generalized hide-and-seek game}
\label{sec:hide_seek}

In this section we show how the LTU matching problem can be reframed as a generalization of the two-person game known as hide-and-seek.
To begin with, we recall the classic hide-and-seek game introduced by \citet{vonNeumann1953}.
In this zero-sum game, player 1 hides in the cell of a matrix while player 2 looks for him, either in a row or in a column.
Reprising our notations from section \ref{sec:LTU_matching}, we index rows by worker type $x \in X$ and columns by job type $y \in Y$.

\paragraph{Classic hide-and-seek.} \begin{enumerate}
\item Player 1 chooses a cell $xy \in X \times Y$ to hide in.
\item Player 2 chooses a row $x' \in X$ or a column $y' \in Y$ to investigate.
\item If player 2 finds player 1 (i.e.\ if $x' = x$ or $y' = y$), then player 1 must pay $\alpha_{xy} > 0$ to player 2; otherwise they both get 0.
\end{enumerate}

\paragraph{}
As shown by von Neumann, finding the Nash equilibria of this game is actually equivalent to solving the assignment problem of workers to jobs:
\begin{equation}
    \max_{\mu \geq 0} ~ \sum_{xy} \mu_{xy} \Phi_{xy}
    \quad \text{s.t. }  (\ref{eq:LTU_stability2}), (\ref{eq:LTU_stability3})
    \label{eq:TU_primal}
\end{equation}
where the surplus $\Phi_{xy}$ produced by assigning worker $x$ to job $y$ is defined as the inverse of the payoff $\alpha_{xy}$, i.e.\ $\Phi_{xy} = 1/\alpha_{xy}$.

In turn, this assignment problem is equivalent to the matching problem between workers and jobs in the TU case \citep{Shapley-Shubik1971}.
Thus any TU matching problem can be reframed as a hide-and-seek game (and vice-versa), and finding the stable outcomes of a TU problem is tantamount to finding the Nash equilibrium strategies of its corresponding game.

This result can be stated as follows:

\begin{theorem*}[\citealt{vonNeumann1953}]
There is a one-to-one mapping between the stable outcomes of the TU matching problem with surpluses $\Phi_{xy}$ and the equilibrium strategies of the classic hide-and-seek game with payoffs $\alpha_{xy} = 1/\Phi_{xy}$.
\end{theorem*}

We will provide the precise nature of this mapping below in (\ref{eq:NE_sol})--(\ref{eq:LTU_sol}), when we extend this result to LTU matching problems.


Let's now consider a non-zero-sum generalization of the classic game.

\paragraph{Generalized hide-and-seek.}
\begin{enumerate}
\item Player 1 chooses a cell $xy \in X \times Y$ to hide in.
\item Player 2 chooses a row $x' \in X$ or a column $y' \in Y$ to investigate.
\item If player 2 finds player 1 by investigating row $x$ (i.e.\ if $x' = x$), then player 1 loses $\alpha_{xy} > 0$ and player 2 wins $\beta_{xy} > 0$.
\item If player 2 finds player 1 by investigating column $y$ (i.e.\ if $y' = y$), then player 1 loses $\gamma_{xy} > 0$ and player 2 wins $\kappa_{xy} > 0$.
\item Otherwise, they both get 0.
\end{enumerate}

\paragraph{}
Note that unlike in classic hide-and-seek, in this generalized version payoffs depend on whether player 2 finds player 1 via a row or via a column.
The classic game is recovered by taking $\alpha_{xy} = \beta_{xy} = \gamma_{xy} = \kappa_{xy}$ for all $x$ and $y$.

We now characterize the Nash equilibria of this generalized hide-and-seek game.
The set of pure strategies for player 1 is $X \times Y$, and for player 2 it is $X \cup Y$.
Their mixed strategy sets are respectively
\begin{align*}
\mathcal S_1
&= \Big\{ p \in \mathbf R^{|X| |Y|}_+ : \sum_{xy \in X \times Y} p_{xy} = 1 \Big\}, \\
\mathcal S_2
&= \Big\{ q \in \mathbf R^{|X| + |Y|}_+ : \sum_{x \in X} q_x + \sum_{y \in Y} q_y = 1 \Big\}.
\end{align*}

Call $\ell(p, q)$ the expected loss of player 1, and $\pi(p, q)$ the expected payoff of player 2.
A pair of mixed strategies $(p^*, q^*)$ is a Nash equilibrium if and only if
\begin{align}
\ell (p^*, q^*) &\leq \ell (p, q^*)
\qquad \, \forall p \in \mathcal S_1, \label{eq:NE_cond1} \\
\pi (p^*, q^*) &\geq \pi (p^*, q)
\qquad \forall q \in \mathcal S_2. \label{eq:NE_cond2}
\end{align}
We will also use a common property of mixed-strategy equilibria, which is that any pure strategy which is played with positive probability in equilibrium must yield the same expected payoff.
Letting $\delta_s$ denote the mixed strategy representation of a pure strategy $s$, we can write this property as follows:
\begin{align}
p^*_{xy} > 0 &\implies \ell (\delta_{xy}, q^*) = \ell (p^*, q^*) \label{eq:NE_cond3} \\
q^*_x > 0 &\implies \pi (p^*, \delta_x) = \pi (p^*, q^*) \label{eq:NE_cond4} \\
q^*_y > 0 &\implies \pi (p^*, \delta_y) = \pi (p^*, q^*) \label{eq:NE_cond5}.
\end{align}

\paragraph{}
We now state our result, which draws the equivalence between the LTU matching problem from section \ref{sec:LTU_matching} and a generalized hide-and-seek game.
In the formulas below, $a^\top b$ denotes the scalar product between two vectors $a$ and $b$.

\begin{theorem}
\label{thm:LTU_HideAndSeek}
There is a one-to-one mapping between the stable outcomes of the LTU matching problem described in section \ref{sec:LTU_matching}, and the equilibrium strategies of the generalized hide-and-seek game with payoffs
\begin{equation}
\alpha_{xy} = \dfrac{\lambda_{xy}}{n_x \Phi_{xy}},
\quad
\beta_{xy} = \dfrac{1}{2 n_x \Phi_{xy}},
\quad
\gamma_{xy} = \dfrac{1-\lambda_{xy}}{m_y \Phi_{xy}},
\quad
\kappa_{xy} = \dfrac{1}{2 m_y \Phi_{xy}}. \label{eq:HS_parametrization}
\end{equation}

Specifically, if $(\mu^*, u^*, v^*)$ is stable for the LTU matching problem, then
\begin{equation}
p^*_{xy} = \frac{\Phi_{xy} \mu^*_{xy}}{\Phi^\top \mu^*},
\quad
q^*_x = \frac{n_x u^*_x}{n^\top u^* + m^\top v^*},
\quad
q^*_y = \frac{m_y v^*_y}{n^\top u^* + m^\top v^*}
\label{eq:NE_sol}
\end{equation}
is a Nash equilibrium of the hide-and-seek game.

Conversely, if $(p^*,q^*)$ is a Nash equilibrium of the hide-and-seek game, then
\begin{equation}
\mu^*_{xy} = \frac{p^*_{xy}}{2 \Phi_{xy} \, \pi(p^*, q^*)},
\quad
u^*_x = \frac{q^*_x}{2 n_x \, \ell(p^*, q^*)}, 
\quad
v^*_y = \frac{q^*_y}{2 m_y \, \ell(p^*, q^*)}
\label{eq:LTU_sol}
\end{equation}
is stable for the LTU matching problem.
\end{theorem}

\paragraph{}
Before delving into the proof, notice that the hide-and-seek game defined by the payoffs (\ref{eq:HS_parametrization}) takes the classic form when the following three conditions are met:
(i) there is one worker and one job per type, i.e.\ $n_x = m_y = 1$;
(ii) we are in the TU case, i.e.\ $\lambda_{xy} = 1/2$ for all $xy$;
and (iii) there are as many workers as there are jobs, i.e.\ $\sum_x n_x = \sum_y m_y = N$.
Indeed, under these conditions we have $\alpha_{xy} = \beta_{xy} = \gamma_{xy} = \kappa_{xy} = 1/\Phi_{xy}$, and our statement reduces to the theorem by von Neumann recalled above.

\begin{proof}
We start with the first implication.
Suppose that $(\mu^*, u^*, v^*)$ is a stable outcome to the LTU matching problem, so that it satisfies the stability conditions (\ref{eq:LTU_stability1})–(\ref{eq:LTU_stability6}).
Because $\Phi_{xy} > 0$ for all $xy$ we must have $\Phi^\top \mu^* > 0$ and $n^\top u^* + m^\top v^* > 0$; if not, any $xy$ would be a blocking pair.
The definition of $p^*$ and $q^*$ in (\ref{eq:NE_sol}) is therefore licit, and it is clear that they indeed constitute a pair of mixed strategies in $\mathcal S_1$ and $\mathcal S_2$ respectively.

Now we show that $p^*$ is a best response to $q^*$ and vice-versa.
For any $p \in \mathcal S_1$, the expected loss of player 1 is
\begin{align*}
\ell(p, q^*)
&= \sum_{xy} p_{xy}
\left( 
\frac{n_x u^*_x}{n^\top u^* + m^\top v^*} \frac{\lambda_{xy}}{n_x \Phi_{xy}}
+ \frac{m_y v^*_y}{n^\top u^* + m^\top v^*} \frac{1-\lambda_{xy}}{m_y \Phi_{xy}}
\right) \\
&= \frac{1}{n^\top u^* + m^\top v^*} \sum_{xy}  p_{xy}
\frac{\lambda_{xy} \, u^*_x + (1-\lambda_{xy}) v^*_y}{\Phi_{xy}}
\geq \frac{1}{2(n^\top u^* + m^\top v^*)},
\end{align*}
where we have used condition (\ref{eq:LTU_stability1}), that is, $\lambda_{xy} \, u^*_x + (1-\lambda_{xy}) \, v^*_y \geq \Phi_{xy}/2$; as well as $\sum_{xy} p_{xy} = 1$.
Furthermore, when $p = p^*$ there is equality in the inequality above.
Indeed, by definition of $p^*_{xy}$ we have that $p^*_{xy} > 0$ implies $\mu^*_{xy} > 0$, which in turn implies $\lambda_{xy} \, u^*_x + (1-\lambda_{xy}) \, v^*_y = \Phi_{xy}/2$ according to (\ref{eq:LTU_stability4}).
Hence $\ell(p, q^*) \geq \ell(p^*, q^*)$ for all $p \in \mathcal S_1$, i.e.\ $p^*$ is a best response to $q^*$.

For player 2, the expected payoff when playing strategy $q \in \mathcal S_2$ is
\begin{align*}
\pi(p^*, q)
&= \sum_{xy} \frac{\Phi_{xy} \mu^*_{xy}}{\Phi^\top \mu^*}
\left( q_x \frac{1}{2 n_x \Phi_{xy}} + q_y \frac{1}{2 m_y \Phi_{xy}} \right) \\
&= \frac{1}{2(\Phi^\top \mu^*)} \sum_{xy} \mu^*_{xy}
\left( q_x \frac{1}{n_x} + q_y \frac{1}{m_y} \right) \\
&= \frac{1}{2(\Phi^\top \mu^*)}
\left( \sum_x q_x \frac{\sum_y \mu^*_{xy}}{n_x} + \sum_y q_y \frac{\sum_x \mu^*_{xy}}{m_y} \right)
\leq \frac{1}{2(\Phi^\top \mu^*)},
\end{align*}
where we have used conditions (\ref{eq:LTU_stability2})--(\ref{eq:LTU_stability3}), that is, $\sum_y \mu^*_{xy} \leq n_x$ and $\sum_x \mu^*_{xy} \leq m_y$; as well as $\sum_x q_x + \sum_y q_y = 1$.
As was the case above with player 1, conditions (\ref{eq:LTU_stability5})--(\ref{eq:LTU_stability6}) now imply that we have equality when $q = q^*$.
Hence $\pi(p^*, q) \leq \pi(p^*, q^*)$ for all $q \in \mathcal S_2$, i.e.\ $q^*$ is a best response to $p^*$.

Thus we have proven that $(p^*, q^*)$ as defined by (\ref{eq:NE_sol}) is a Nash equilibrium of the generalized hide-and-seek game.

Now we prove the converse.
Suppose that $(p^*, q^*)$ is a Nash equilibrium of the generalized hide-and-seek game.
The equilibrium payoff of player 2, $\pi(p^*, q^*)$, must be positive: if not, playing any pure strategy $x$ such that $p^*_{xy} > 0$ for some $y$ would be a profitable deviation.
Furthermore, because player 1 incurs a positive loss whenever player 2 gets a positive payoff, we must also have $\ell(p^*, q^*) > 0$.
Thus the definitions of $\mu^*$, $u^*$ and $v^*$ in (\ref{eq:LTU_sol}) are licit.

We want to prove that $(\mu^*, u^*, v^*)$ so defined verifies the stability conditions (\ref{eq:LTU_stability1})–(\ref{eq:LTU_stability6}).
First we write the best response condition (\ref{eq:NE_cond1}) for $\delta_{xy} \in \mathcal S_1$, the pure strategy $xy$ of player 1:
\begin{equation*}
\ell (p^*, q^*) \leq \ell (\delta_{xy}, q^*)
= q^*_x \frac{\lambda_{xy}}{n_x \Phi_{xy}} + q^*_y \frac{1-\lambda_{xy}}{m_y \Phi_{xy}}.
\end{equation*}
Multiplying by $\Phi_{xy}/2 \ell (p^*, q^*)$ and using the definitions of $u^*_x$ and $v^*_y$, we obtain
\begin{equation*}
\Phi_{xy}/2 \leq \lambda_{xy} \, u^*_x + (1-\lambda_{xy}) \, v^*_y,
\end{equation*}
i.e.\ stability condition $(\ref{eq:LTU_stability1})$.
Furthermore, if $\mu^*_{xy} > 0$ then $p^*_{xy} > 0$, which by (\ref{eq:NE_cond3}) implies equality in the inequality above.
Thus $(\mu^*, u^*, v^*)$ also satisfies $(\ref{eq:LTU_stability4})$.

Next we write the best response condition (\ref{eq:NE_cond2}) for the pure strategy $x$ of player 2:
\begin{equation*}
\pi (p^*, q^*) \geq \pi (p^*, \delta_x)
= \sum_y p^*_{xy} \frac{1}{2 n_x \Phi_{xy}}.
\end{equation*}
Multiplying by $n_x / \pi (p^*, q^*)$ and using the definition of $\mu^*_{xy}$ we obtain (\ref{eq:LTU_stability2}),
\begin{equation*}
n_x \geq \textstyle \sum_y \mu^*_{xy}.
\end{equation*}
As before, if $u^*_x > 0$ then $q^*_x > 0$, which by (\ref{eq:NE_cond4}) implies equality in the inequality above.
Thus $(\mu^*, u^*, v^*)$ also satisfies $(\ref{eq:LTU_stability5})$.

Finally, by writing the best response condition (\ref{eq:NE_cond2}) for the pure strategy $y$ of player 2, and by using (\ref{eq:NE_cond5}), we similarly obtain that $(\mu^*, u^*, v^*)$ satisfies the stability conditions $(\ref{eq:LTU_stability3})$ and $(\ref{eq:LTU_stability6})$.

Thus we have shown that the outcome $(\mu^*, u^*, v^*)$ defined by (\ref{eq:LTU_sol}) is stable for the LTU matching problem.

Since $\pi(p^*, q^*) = \frac{1}{2(\Phi^\top \mu^*)}$ and $\ell(p^*, q^*) = \frac{1}{2(n^\top u^* + m^\top v^*)}$, the formulas (\ref{eq:NE_sol}) and (\ref{eq:LTU_sol}) are inverse of each other,
and therefore they constitute a one-to-one mapping between the two sets of solutions.
\end{proof}

\paragraph{}
It is worth noting that even though we have focused our discussion on the more common case of one-to-one matching, theorem 1 can actually be adapted to encompass problems of many-to-one matching with linear transfers as well.
(This also includes the roommate problem.)
In appendix \ref{apx:many-to-one} we generalize the matching model presented in section \ref{sec:LTU_matching} by allowing for a population of individuals to match however they want between themselves, in arrangements of up to $N$ members.
We show that this more general matching problem is equivalent to a $N$-dimensional version of the hide-and-seek game, in the sense that the game is played in an $N$-dimensional array.
This generalization also shows how to handle the case of nonpositive outputs.

\paragraph{}
With theorem \ref{thm:LTU_HideAndSeek} (and its many-to-one generalization) we thus prove that any LTU matching problem can be reframed as a two-person game.
In addition, we establish a correspondence between the stable outcomes of the former and the Nash equilibria of the latter.
In doing so, we generalize the link between TU matching problems and the classic hide-and-seek game studied by \citeauthor{vonNeumann1953}.

We see two important implications to this result.
The first implication is a new proof for the existence of stable outcomes in LTU matching problems, which follows immediately from our result via the existence of Nash equilibria in any finite two-person game \citep{Nash1950}.

\begin{corollary}
Any LTU matching problem with a finite number of types and bounded-size arrangements has a stable outcome.
\end{corollary}


\paragraph{}
This provides an alternative route to previous existence results in two-sided matching, such as \citet{Quinzii1984}, \citet{Alkan-Gale1990}, or more recently \citet{Noldeke-Samuelson2018} (all three covering the more general ITU case), to matching problems involving any bounded-size arrangements between individuals.
Additionally, our approach uniquely connects this class of matching problems with game theory.



\paragraph{}
The second consequence pertains to the computation of solutions to the matching problem.
Since we have shown that solving the LTU matching problem is equivalent to solving a hide-and-seek game, standard algorithms used to compute Nash equilibria of two-person games (e.g.\ Lemke–Howson) can be used to retrieve stable outcomes (see \citealt{vonStengel2002} for an exhaustive review).
This observation opens up many alternatives to compute stable outcomes in the LTU case, in addition to existing methods for the more general ITU case (e.g.\ \citealt{Alkan1989}).



\section{LTU and linear programming}
\label{sec:exchangeability_TU}


A fundamental property of matching problems with TU is that any stable outcome maximizes the total surplus -- hence the link with the assignment problem \citep{Shapley-Shubik1971}.
A natural question which arises is thus whether this property carries over to the LTU case or not.
The first difficulty, as discussed in section \ref{sec:LTU_matching}, is that the joint output of the match $\Phi_{xy}$ cannot be interpreted as surplus anymore in the LTU case, since it holds different value for the worker and the employer.
Still, we could imagine that an assignment which maximizes the total output would leave more utility to be shared between the agents, and therefore lead to a stable outcome.

In this section we show that this intuition actually breaks down for LTU problems.
More specifically, LTU problems cannot be reframed as linear programs unless they belong to the subclass of TU problems.
This means in particular that the total output typically varies across stable outcomes in LTU problems, unlike in TU problems.

\paragraph{}
To obtain this result, we investigate a remarkable property of TU matching problems and zero-sum games, which is that their solutions are \emph{exchangeable}.
For instance in the case of a TU problem, it means that if $(\mu, u, v)$ and $(\mu', u', v')$ are any two stable outcomes, then $(\mu', u, v)$ is also a stable outcome.
This exchangeability can be seen as a consequence of their underlying structure as linear programs \citep{Dantzig1951}.

This property is however not inherited by LTU matching models in general, as shown by the following example.

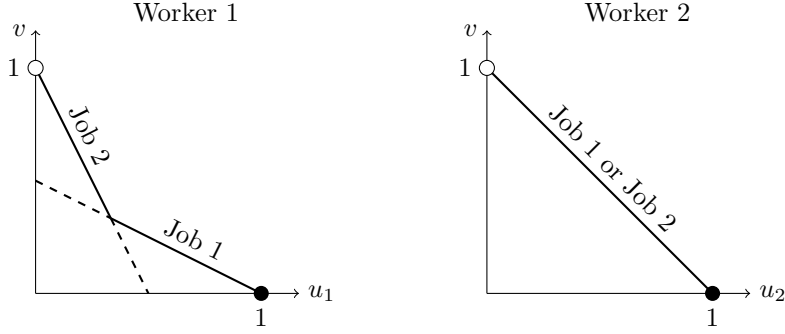
\begin{figure}[t]
\centering
\begin{tikzpicture}
    \node[above] at (2,3.5) {Worker 1};
    \draw[->] (0,0) -- (3.5,0) node[right] {$u_1$};
    \draw[->] (0,0) -- (0,3.5) node[left] {$v$};
    \draw[thick, dashed] (0,1.5) -- (1,1);
    \draw[thick] (1,1) -- (3,0)
        node[pos=1, below, yshift=-2pt] {1}
        node[pos=0.5, sloped, above] {Job 1};
    \draw[thick] (0,3) -- (1,1)
        node[pos=0, left, xshift=-2pt] {1}
        node[pos=0.5, sloped, above] {Job 2};
    \draw[thick, dashed] (1,1) -- (1.5,0);
    \fill (3,0) circle (.1);
    \draw[fill=white] (0,3) circle (.1);

    \node[above] at (8,3.5) {Worker 2};
    \draw[->] (6,0) -- (9.5,0) node[right] {$u_2$};
    \draw[->] (6,0) -- (6,3.5) node[left] {$v$};
    \draw[thick] (6,3) -- (9,0)
        node[pos=0, left, xshift=-2pt] {1}
        node[pos=1, below, yshift=-2pt] {1}
        node[pos=0.5, sloped, above] {Job 1 or Job 2};
    \fill (9,0) circle (.1);
    \draw[fill=white] (6,3) circle (.1);
\end{tikzpicture}
\caption{Example of a LTU problem with non-exchangeable solutions.}
\label{fig:nonexchangeability}
\end{figure}

\paragraph{\textnormal{\itshape Example.}}
Suppose there are two workers ($x = 1, 2$) and two jobs ($y = 1, 2$).
The within-match feasibility constraints on utilities are
\begin{equation*}
u_1 + 2 v_1 = 1, \qquad
2 u_1 + v_2 = 1, \qquad
u_2 + v_1 = 1, \qquad
u_2 + v_2 = 1.
\end{equation*}
They are illustrated on figure \ref{fig:nonexchangeability}.
One can verify that the two following outcomes are stable:
$\mu_{11} = \mu_{22} = 1$, $u_1 = u_2 = 1$, $v_1 = v_2 = 0$ (black dots in figure \ref{fig:nonexchangeability});
and $\mu_{12}' = \mu_{21}' = 1$, $u_1' = u_2' = 0$, $v_1' = v_2' = 1$ (white dots in figure \ref{fig:nonexchangeability}).
Yet, the outcome $(\mu', u, v)$ is not stable – in fact it is not feasible, since worker 1 cannot reach a utility of 1 when matched with job 2 (in figure \ref{fig:nonexchangeability}, the black dot is outside their feasible set).




\paragraph{}
We will show that the exchangeability of solutions actually characterizes TU problems among LTU problems.
We start by defining this subclass.

\begin{definition}
A LTU matching problem has the TU property if there exist positive numbers $a_x$ for $x \in X$ and $b_y$ for $y \in Y$ such that, by rescaling utilities as $\tilde u_x = a_x u_x$ and $\tilde v_y = b_y v_y$, the within-match utility constraints (\ref{eq:LTU}) rewrite as
\begin{equation*}
\tilde u_x + \tilde v_y = \tilde \Phi_{xy}
\end{equation*}
for some numbers $\tilde \Phi_{xy}$. (If this is the case, then $\tilde \Phi_{xy} = (a_x + b_y) \Phi_{xy}/2$.)
\end{definition}

Here we remark that any LTU problem with $|X| = 1$ has the TU property:
one can take for instance $a_1 = 1$ and $b_y = (1-\lambda_{1y})/\lambda_{1y}$.
(By symmetry, this is also true when $|Y| = 1$.)

Our goal is to link the subclass of TU problems with the concept of solution exchangeability.
Yet, some LTU problems do not have the TU property, but do have exchangeable solutions.
For instance, when $\Phi_{xy} = 0$ for all $xy$, and irrespective of the coefficients $\lambda_{xy}$, any feasible $\mu$ with $u_x = v_y = 0$ constitutes a stable outcome.
The TU property is therefore not equivalent with having exchangeable solutions, so we must specify the exact relationship between solution exchangeability and the TU property.

This relationship will rely on the notion of \emph{subproblem}, which we introduce here.
In section \ref{sec:LTU_matching}, we defined a LTU matching problem by (i) its type sets $X$ and $Y$, (ii) the masses of individuals of each type $n_x$ and $m_y$, (iii) the numbers $\lambda_{xy}$ and $\Phi_{xy}$ specifying its within-match feasibility constraints on utility, and (iv) the reservation utilities, henceforth denoted $\underline u_x$ and $\underline v_y$, which were set to 0 by default.
Overall, this LTU problem can thus be denoted $(X, Y, n, m, \lambda, \Phi, \underline u, \underline v)$.
We then define a subproblem as follows.

\begin{definition}
A subproblem of the LTU problem $(X, Y, n, m, \lambda, \Phi, \underline u, \underline v)$ is any LTU problem $(\tilde X, \tilde Y, \tilde n, \tilde m, \tilde \lambda, \tilde \Phi, \underline{\tilde u}, \underline{\tilde v})$ such that $\tilde X \subset X$, $\tilde Y \subset Y$, and for all $x \in \tilde X$ and $y \in \tilde Y$, $\tilde \lambda_{xy} = \lambda_{xy}$ and $\tilde \Phi_{xy} = \Phi_{xy}$.
\end{definition}

In other words, a subproblem is obtained by keeping intact the within-match feasibility constraints (\ref{eq:LTU}) (i.e.\ the numbers $\lambda_{xy}$ and $\Phi_{xy}$), but possibly changing the mass of individuals of each type (or even removing a type entirely) as well as the reservation utilities.
This definition could also require that $\tilde n_x \leq n_x$ and $\tilde m_y \leq m_y$ – hence the name \emph{subproblem}, since this would constitute a subset of the original matching market.
However, ignoring this requirement is without loss of generality since we can always rescale the population mass with no consequence on stability.

We then say that a LTU problem has totally exchangeable solutions when all its subproblems have exchangeable solutions.


\begin{definition}
A LTU matching problem has totally exchangeable solutions if all its subproblems are such that, if $(\mu, u, v)$ and $(\mu', u', v')$ are two stable outcomes to this subproblem, then $(\mu', u, v)$ is also stable for this subproblem.
\end{definition}

With this, we are finally able to state our result.

\begin{theorem}
\label{thm:nonexchangeability}
A LTU matching problem has the TU property if and only if it has totally exchangeable solutions.
\end{theorem}

The proof is found in Appendix \ref{app:exchangeability_proofs}.
The direct implication simply follows from the exchangeability of solutions in TU problems, since their subproblems are still TU problems.
The difficulty lies therefore with the reverse implication.
To prove it, we show that when a LTU problem does not have the TU property, we can construct a subproblem with $2 \times 2$ types which has two non-exchangeable solutions.

Since all linear programming problems have exchangeable solutions, we are able to conclude by stating that among LTU problems, only TU problems can be formulated as linear programs.

\begin{corollary}
A LTU matching problem is a linear program if and only if it has the TU property.
\end{corollary}

This last result cements the tight relationship between TU matching problems, zero-sum hide-and-seek games, and linear programs that was initially investigated by \citeauthor{vonNeumann1953}.
It shows that even a slight variation on the TU paradigm, which is to keep utility transfers linear but relax the one-for-one transfer rate assumption, is enough to move the matching problem outside the scope of linear programming.

\paragraph{}
Overall, the results presented in this note extend the previously-known relationship between TU matching problems and zero-sum hide-and-seek games, by looking at the non-TU case.
While the linear programming structure is lost when moving away from the TU case, matching problems with LTU retain a convenient equivalence with the hide-and-seek game, although in a bimatrix (i.e.\ nonzero-sum) version.
This equivalence notably provides a practical way to compute stable outcomes to LTU matching problems via the methods which exist to find Nash equilibria in bimatrix games.




\bibliographystyle{agsm}
\bibliography{references}

@article{vonNeumann1953,
  title={A certain zero-sum two-person game equivalent to the optimal assignment problem},
  author={von Neumann, John},
  journal={Contributions to the Theory of Games},
  volume={2},
  pages={5--12},
  year={1953}
}

@article{Nash1950,
  title={Equilibrium points in {$N$}-person games},
  author={Nash, John},
  journal={Proceedings of the National Academy of Sciences of the United States of America},
  volume={36},
  number={1},
  pages={48--49},
  year={1950}
}

@article{Crawford-Knoer1981,
  title={Job matching with heterogeneous firms and workers},
  author={Vincent P. Crawford and Elsie Marie Knoer},
  journal={Econometrica},
  volume={49},
  number={2},
  pages={437--450},
  year={1981}
}

@article{Noldeke-Samuelson2018,
  title={The implementation duality},
  author={Georg Nöldeke and Larry Samuelson},
  journal={Econometrica},
  volume={86},
  number={4},
  pages={1283--324},
  year={2018}
}

@article{Galichon-Kominers-Weber2019,
  title={Costly concessions: An empirical framework for matching with imperfectly transferable utility},
  author={Galichon, A. and Kominers, S. D. and Weber, S.},
  journal={Journal of Political Economy},
  volume={127},
  number={6},
  pages={2875--2925},
  year={2019}
}

@article{Kelso-Crawford1982,
  title={Job matching, coalition formation, and gross substitutes},
  author={Alexander S. Kelso and Vincent P. Crawford},
  journal={Econometrica},
  volume={50},
  number={6},
  pages={1483--1504},
  year={1982}
}

@book{Roth-Sotomayor1990,
  title={Two-sided matching: A study in game-theoretic modeling and analysis},
  author={Alvin E. Roth and Marilda A. Oliveira Sotomayor},
  year={1990},
  publisher={Cambridge University Press},
  number={18},
  series={Econometric Society Monographs}
}

@article{Alkan-Gale1990,
  title={The core of the matching game},
  author={Ahmet Alkan and David Gale},
  journal={Games and Economic Behavior},
  volume={2},
  number={3},
  pages={203--212},
  year={1990}
}

@article{Alkan1989,
  title={Existence and computation of matching equilibria},
  author={Ahmet Alkan},
  journal={European Journal of Political Economy},
  volume={5},
  number={2--3},
  pages={285--296},
  year={1989}
}

@article{Quinzii1984,
  title={Core and competitive equilibria with indivisibilities},
  author={Martine Quinzii},
  journal={International Journal of Game Theory},
  volume={13},
  number={1},
  pages={41--60},
  year={1984}
}

@article{Shapley-Shubik1971,
  title={The assignment game {I}: the core},
  author={L. S. Shapley and M. Shubik},
  journal={International Journal of Game Theory},
  volume={1},
  number={},
  pages={111--130},
  year={1971}
}

@incollection{Dantzig1951,
    author={G. B. Dantzig},
    title={A proof of the equivalence of the programming problem and the game problem},
    booktitle={Activity Analysis of Production and Allocation},
    editor={T. C. Koopmans},
    series={Cowles Commission Monographs},
    number={13},
    pages={330--335},
    publisher={Wiley},
    address={New York},
    year={1951}
}

@article{vonStengel2023,
  title={Zero-Sum Games and Linear Programming Duality},
  author={Bernhard von Stengel},
  journal={Mathematics of Operations Research},
  volume={},
  number={},
  pages={1-18},
  year={2023}
}

@article{vonStengel2002,
  title={Computing equilibria for two-person games},
  author={Bernhard von Stengel},
  journal={Handbook of game theory with economic applications},
  volume={3},
  number={},
  pages={1723--1759},
  year={2002}
}

@article{Scarf1967,
  title={The Core of an {N} Person Game},
  author={Herbert E. Scarf},
  journal={Econometrica},
  volume={35},
  number={1},
  pages={50--69},
  year={1967}
}

@article{Adler2013,
  title={The equivalence of linear programs and zero-sum games},
  author={I. Adler},
  journal={International Journal of Game Theory},
  volume={42},
  number={},
  pages={165--177},
  year={2013}
}

@article{Hatfield-Milgrom2005,
  title={Matching with contracts},
  author={John William Hatfield and Paul R. Milgrom},
  journal={American Economic Review},
  volume={95},
  number={4},
  pages={913--935},
  year={2005}
}

@book{Chiappori2017,
  author = {Pierre-Andr{\'e} Chiappori},
  publisher = {Princeton University Press},
  title = {Matching with transfers: The economics of love and marriage},
  year = {2017}
}

@article{Chiappori-Reny2016,
  title={Matching to share risk},
  author = {Pierre-Andr{\'e} Chiappori and Philip J. Reny},
  journal={Theoretical Economics},
  volume={11},
  number={1},
  pages={227--251},
  year={2016}
}

@article{Legros-Newman2007,
  title={Beauty is a beast, frog is a prince: Assortative matching with nontransferabilities},
  author = {Patrick Legros and Andrew F. Newman},
  journal={Econometrica},
  volume={75},
  number={4},
  pages={1073--1102},
  year={2007}
}

\newpage
\appendix

\section*{\huge Appendix}

\section{Many-to-one matching with LTU}
\label{apx:many-to-one}


Here we show how theorem \ref{thm:LTU_HideAndSeek} can be extended to the larger class of many-to-one matching problems with linearly transferable utility.
We allow for a population of individuals to match however they want between themselves, in arrangements of up to $N$ members.
This is a many-to-one matching problem because we can consider that all individuals to be matched are on one side of the market, and on the other side are labels corresponding to each possible arrangement.
We show that this matching problem is equivalent to a $N$-dimensional version of the hide-and-seek game.

We consider a population of individuals who belong to a finite number of types $x \in X$.
The mass of individuals of type $x$ is $n_x$.
Individuals are able to form \emph{arrangements} of up to $N$ members, which are defined as $N$-tuples
\begin{equation}
    a = (x_1, x_2, \dots, x_N) \in (X \cup \{0\})^N.
\end{equation}
Thus each spot $i$ in the arrangement is either taken by an individual, $x_i \in X$, or left vacant, $x_i = 0$.
We denote by $M_{x, a}$ the number of individuals with type $x$ in arrangement $a$.

Note that under this definition of an arrangement, placement can matter.
For instance, $(x, x', \dots, 0)$ can \emph{a priori} be different from $(x', x, \dots, 0)$.
(As an illustration, consider how a football team might fare if its goalkeeper were to switch positions with a striker.)
In particular, for any individual $x$ there are $N$ distinct arrangements where $x$ is single, namely $(x, 0, \dots, 0)$, $(0, x, \dots, 0)$, \dots, $(0, 0, \dots, x)$.
If so desired however, we can easily consider that placement does not matter by excluding all redundant arrangements.

In fact, we will consider that only a subset $A \subset (X \cup \{0\})^N$ from all potential arrangements are allowed.
We need make only two restrictions on this set $A$: first, it must exclude the empty arrangement $(0, 0, \dots, 0)$ (this is purely for convenience); and second, it must include for all types $x$ at least one arrangement whereby $x$ is single.

We now extend the notion of linearly transferable utility to arrangements.
Let $u_x$ denote the utility of individuals of type $x$.
We assume that for any arrangement $a \in A$, there are $N$ nonnegative numbers $\lambda_{i, a}$, the sum of which is 1, and verifying $\lambda_{i, a} = 0$ if and only if $x_i = 0$; and a number $\Phi_a$, such that the utility feasibility constraint within arrangement $a$ is
\begin{equation}
    \label{eq:LTU_arr}
    \textstyle \sum_i \lambda_{i, a} \, u_{x_i} = \Phi_a.
\end{equation}
Note that when $a$ is made of a single individual of type $x$, (\ref{eq:LTU_arr}) writes as $u_x = \Phi_a$ and $\Phi_a$ is therefore a reservation utility for this type.
Since $\sum_i \lambda_{i,a} = 1$, we can thus add any constant $K$ to all outputs $\Phi_a$ without changing the nature of the problem.
For this reason, we can assume without loss of generality that $\Phi_a > 0$ for all $a$.
(We could not do this in section \ref{sec:LTU_matching} because reservation utilities were fixed at 0.)

A matching $\mu$ collects the masses $\mu_a$ of arrangements $a \in A$.  It is feasible if
\begin{equation}
    \label{eq:feasible_matching_arr}
    \textstyle \sum_a M_{x, a} \, \mu_a = n_x
    \qquad \forall x \in X.
\end{equation}
An outcome $(\mu, u)$ is stable if its matching $\mu$ is feasible and
\begin{align}
    \textstyle \sum_i \lambda_{i, a} \, u_{x_i} &\geq \Phi_a
    \qquad \forall a \in A \\
    \mu_a > 0 &\implies \textstyle \sum_i \lambda_{i, a} \, u_{x_i} = \Phi_a.
\end{align}
This concludes our presentation of the general LTU matching problem.

\paragraph{}
Such problems are in fact still equivalent to a hide-and-seek game, albeit now played in an $N$-dimensional array.
Each dimension of this array is indexed by $x \in X \cup \{0\}$.
Player 1 hides in a cell $a \in A$, and player 2 looks for him by choosing an index $x$.
We say that player 2 finds player 1 if one of the indices of player 1's hiding cell is $x$.

\paragraph{$N$-dimensional hide-and-seek.}
\begin{enumerate}
\item Player 1 chooses a cell $a \in A$ to hide in.
\item Player 2 chooses an index $x \in X$ to investigate.
\item If player 2 finds player 1 (i.e.\ if $M_{x,a} \geq 1$), then player 1 loses $\alpha_{x, a} > 0$ and player 2 wins $\beta_{x, a} > 0$; otherwise they both get 0.
\end{enumerate}

\paragraph{}
The following result extends theorem \ref{thm:LTU_HideAndSeek} to the more general class of LTU matching problems introduced in this section, in which individuals can match in arbitrary arrangements of up to $N$ members.

\begin{theorem}
\label{thm:LTU_HideAndSeek_Ndim}
There is a one-to-one mapping between the stable outcomes of the many-to-one LTU matching problem described above, and the equilibrium strategies of the $N$-dimensional hide-and-seek game with payoffs
\begin{equation}
\alpha_{x,a} = \frac{\sum_{i | x_i = x} \lambda_{i, a}}{n_x \Phi_a},
\qquad
\beta_{x,a} = \frac{M_{x, a}}{n_x \Phi_a}.
\end{equation}

Specifically, if $(\mu^*, u^*)$ is stable for the LTU matching problem, then
\begin{equation}
p^*_a = \frac{\Phi_a \mu^*_a}{\Phi^\top \mu^*},
\qquad
q^*_x = \frac{n_x u^*_x}{n^\top u^*}
\end{equation}
is a Nash equilibrium of the hide-and-seek game.

Conversely, if $(p^*,q^*)$ is a Nash equilibrium of the hide-and-seek game, then
\begin{equation}
\mu^*_a = \frac{p^*_a}{\Phi_a \, \pi(p^*, q^*)}
\qquad
u^*_x = \frac{q^*_x}{n_x \, \ell(p^*, q^*)}
\end{equation}
is stable for the LTU matching problem.
\end{theorem}



\paragraph{}
\begin{proof}
We start with the first implication.
Suppose that $(\mu^*, u^*)$ is a stable outcome for the LTU matching problem.
Since $\Phi_a > 0$ for all $a \in A$ by assumption, and there must be at least one positive $\mu^*_a$, we have $\Phi^\top \mu^* > 0$ and the definition of $p^*_a$ is licit.
Similarly, $u_x^* > 0$ for all $x$ (otherwise $x$ could do better by remaining single) so $n^\top u^* > 0$ and the definition of $q^*_x$ is also licit.

We now compute the expected payoffs using the definitions of $p^*$ and $q^*$.
For player 1,
\begin{align*}
\ell(p^*, q^*)
&= \sum_a \frac{\Phi_a \mu^*_a}{\Phi^\top \mu^*}
\left( 
\sum_x \frac{n_x u^*_x}{n^\top u^*} \frac{\sum_{i | x_i = x} \lambda_{i, a}}{n_x \Phi_a}
\right) \\
&= \frac{1}{(\Phi^\top \mu^*) \, (n^\top u^*)} \sum_a \mu^*_a \sum_x 
\left(
\sum_{i | x_i = x} \lambda_{i, a}
\right) u^*_x \\
&= \frac{1}{(\Phi^\top \mu^*) \, (n^\top u^*)} \sum_a \mu^*_a \sum_i \lambda_{i, a} \, u^*_{x_i} \\
&= \frac{1}{(\Phi^\top \mu^*) \, (n^\top u^*)} \sum_a \mu^*_a \Phi_a
= \frac{1}{m^\top u^*},
\end{align*}
where we have used that $\mu^*_a > 0$ implies $\sum_i \lambda_{i, a} \, u^*_{x_i} = \Phi_a$.
In addition, for any $p \in \mathcal S_1$ we have
\begin{align*}
\ell(p, q^*)
&= \sum_a p_a
\left( 
\sum_x \frac{n_x u^*_x}{m^\top u^*} \frac{\sum_{i | x_i = x} \lambda_{i, a}}{n_x \Phi_a}
\right) \\
&= \frac{1}{m^\top u^*} \sum_a p_a \frac{\sum_n \lambda_{i,a} \, u^*_{x_i}}{\Phi_a} \\
&\geq \frac{1}{m^\top u^*} \sum_a p_a
= \frac{1}{m^\top u^*} = \ell(p^*, q^*),
\end{align*}
where this time we have used that $\sum_i \lambda_{i,a} \, u^*_{x_i} \geq \Phi_a$, as well as $\sum_a p_a = 1$.
Hence $p^*$ is a best response to $q^*$.

Now, the expected payoff for player 2 when playing any strategy $q \in \mathcal S_2$ is
\begin{align*}
\pi(p^*, q)
&= \sum_a \frac{\Phi_a \mu^*_a}{\Phi^\top \mu^*}
\left( 
\sum_x q_x \frac{M_{x, a}}{n_x \Phi_a}
\right) \\
&= \frac{1}{\Phi^\top \mu^*} \sum_a \mu^*_a \sum_x q_x \frac{M_{x, a}}{n_x} \\
&= \frac{1}{\Phi^\top \mu^*} \sum_x q_x \frac{\sum_a M_{x, a} \, \mu^*_a}{n_x} \\
&= \frac{1}{\Phi^\top \mu^*} \sum_x q_x
= \frac{1}{\Phi^\top \mu^*}
\end{align*}
where we have used $\sum_a M_{x, a} \, \mu^*_a = n_x$ and $\sum_x q_x = 1$.
Here, player 2 is actually indifferent between all strategies against $p^*$, so in particular $q^*$ is a best response to $p^*$.
Thus $(p^*, q^*)$ is a Nash equilibrium of the $N$-dimensional hide-and-seek game.

\paragraph{}
We now prove the converse.
Suppose that $(p^*, q^*)$ is a Nash equilibrium of the $N$-dimensional hide-and-seek game.
The equilibrium payoff of player 2, $\pi(p^*, q^*)$, must be positive: otherwise, player 2 could play any type $x$ belonging to an arrangement which is played with positive probability in $p^*$, for a profitable deviation.
Furthermore, because player 1 incurs a positive loss whenever player 2 gets a positive payoff, we must also have $\ell(p^*, q^*) > 0$.
Thus the definitions of $\mu^*$ and $u^*$ are licit.

We want to show that $(\mu^*, u^*)$ satisfies the feasibility and stability conditions.
Let's write the best response condition for the pure strategy $a$ of player 1:
\begin{equation*}
\ell (p^*, q^*) \leq \ell (\delta_a, q^*)
= \sum_x q^*_x \frac{\sum_{i | x_i = x} \lambda_{i, a}}{n_x \Phi_a}
= \sum_i \lambda_{i, a} \frac{q^*_{x_i}}{n_{x_i} \Phi_a}.
\end{equation*}
Multiplying by $\Phi_a / \ell (p^*, q^*)$ and using the definition of $u^*_x$ we obtain
\begin{equation*}
\Phi_a
\leq \sum_i \lambda_{i, a} \frac{q^*_{x_i}}{n_{x_i} \ell (p^*, q^*)}
= \sum_i \lambda_{i, a} \, u^*_{x_i},
\end{equation*}
which is the first stability condition.
Furthermore, if $\mu^*_a > 0$ then $p^*_a > 0$, which implies equality in the inequality above.
Thus $(\mu^*, u^*)$ also satisfies the second stability condition.

Now, note that at a Nash equilibrium player 2 must be indifferent between all pure strategies $x \in X$, and thus they must all yield the same payoff $\pi(p^*, q^*)$.
Indeed, if there were some $x$ such that $\pi(p^*, \delta_x) < \pi(p^*, q^*)$, then we should have $q^*_x = 0$.
But in this case, player 1 would have a strict profitable deviation with choosing a pure strategy $a$ in which $x$ is single, yielding a loss of $0 < \ell (p^*, q^*)$.
Hence for all $x$,
\begin{equation*}
\pi (p^*, q^*) = \pi (p^*, \delta_x)
= \sum_a p^*_a \frac{M_{x, a}}{n_x \Phi_a}.
\end{equation*}
Multiplying by $n_x / \pi (p^*, q^*)$ and using the definition of $\mu^*_a$ we obtain
\begin{equation*}
n_x = \sum_a M_{x, a} \frac{p^*_a}{\Phi_a \pi (p^*, q^*)}
= \sum_a M_{x,a} \, \mu^*_a
\end{equation*}
and therefore $\mu^*$ is feasible.

Thus we have shown that the outcome $(\mu^*, u^*)$ is stable for the LTU matching problem.
\end{proof}

\newpage

\section{Proof of theorem \ref{thm:nonexchangeability}}
\label{app:exchangeability_proofs}

We start with a proposition, which provides a characterization of the TU property based on the coefficients $\lambda_{xy}$ of the LTU problem.
To keep equations more concise, we introduce
\begin{equation*}
\omega_{xy} \overset{\text{def}}{=} \frac{\lambda_{xy}}{1-\lambda_{xy}}.
\end{equation*}

\begin{proposition}
A LTU matching problem has the TU property if and only if
for all $x, x' \in X$ and all $y, y' \in Y$,
\begin{equation}
\frac{\omega_{xy} \; \omega_{x'y'}}{\omega_{x'y} \; \omega_{xy'}} = 1.
\label{eq:TU_property}
\end{equation}
\end{proposition}

\begin{proof}
Suppose that the problem has the TU property.
Then the constraint $xy$ can be rewritten as $a_x u_x + b_y v_y = (a_x + b_y) \Phi_{xy} /2$.
Dividing by $a_x + b_y$, we identify $\lambda_{xy} = a_x/(a_x + b_y)$ and $1-\lambda_{xy} = b_y/(a_x + b_y)$, so that $\omega_{xy} = \lambda_{xy} / (1-\lambda_{xy}) = a_x/b_y$.
Hence
\begin{equation*}
\frac{\omega_{xy} \; \omega_{x'y'}}{\omega_{x'y} \; \omega_{xy'}}
= \frac{a_x}{b_y} \frac{a_{x'}}{b_{y'}} \frac{b_y}{a_{x'}} \frac{b_{y'}}{a_x}
= 1.
\end{equation*}

Conversely, suppose that this ratio is 1 for all $x, x', y, y'$.
Choose some $x_0 \in X$ and $y_0 \in Y$ arbitrarily, and define
$a_x = \omega_{x y_0}/\omega_{x_0 y_0}$ for all $x$, and $b_y = 1/\omega_{x_0 y}$ for all $y$.
Divide the $xy$ constraint by $(1-\lambda_{xy}) \omega_{x_0 y}$ to get
\begin{equation*}
\underbrace{\frac{\omega_{xy}}{\omega_{x_0 y}} \frac{\omega_{x_0 y_0}}{\omega_{x y_0}}}_{=1} a_x u_x + b_y v_y
= \frac{\Phi_{xy}}{2(1-\lambda_{xy})\omega_{x_0 y}}
\end{equation*}
and therefore the problem has the TU property.
\end{proof}

Now recall the statement to prove:

\begin{theorem*}
A LTU matching problem has the TU property if and only if it has totally exchangeable solutions,
i.e.\ if for any subproblem, and any two solutions to this subproblem, these two solutions are exchangeable.
\end{theorem*}

As mentioned in the main text, the direct implication is immediate, hence we focus on proving the reverse implication.

\begin{proof}
Suppose that a LTU problem does not have the TU property (this notably implies $|X| \geq 2$ and $|Y| \geq 2$, as we remark in the main text).
By our proposition above, it means that there are some types $x, x' \in X$ and $y, y' \in Y$ such that (\ref{eq:TU_property}) does not hold.
(Necessarily, $x \neq x'$ and $y \neq y'$.)
For convenience, we label them as $x = y = 1$ and $x' = y' = 2$.
We have
\begin{equation*}
\rho \overset{\text{def}}{=} \frac{\omega_{11} \; \omega_{22}}{\omega_{21} \; \omega_{12}} \neq 1.
\end{equation*}  
In what follows, we study the case whereby this ratio $\rho$ is \emph{positive}.
(The case $\rho$ negative is solved in a symmetric fashion.)

Consider a subproblem such that $\tilde n_1 = \tilde n_2 = \tilde m_1 = \tilde m_2 = 1$.
Rescale utilities according to
\begin{equation*}
\tilde u_1 = \frac{\omega_{11}}{\omega_{21}} u_1, \qquad
\tilde u_2 = u_2, \qquad
\tilde v_1 = \frac{1}{\omega_{21}} v_1, \qquad
\tilde v_2 = \frac{1}{\omega_{22}} v_2.
\end{equation*}
With these rescalings, the within-match feasibility constraints rewrite as
\begin{align*}
\tilde u_1 + \tilde v_1 &= \tilde \Phi_{11} \\
(1/\rho) \, \tilde u_1 + \tilde v_2 &= \tilde \Phi_{12} \\
\tilde u_2 + \tilde v_1 &= \tilde \Phi_{21} \\
\tilde u_2 + \tilde v_1 &= \tilde \Phi_{22}.
\end{align*}

Now, notice that changing the reservation utilities is equivalent to keeping them at 0 but modifying the outputs in a certain way.
For instance, setting $\underline{\tilde u}_1 = -1$ is equivalent to increasing $\tilde \Phi_{11}$ by 1 and $\tilde \Phi_{12}$ by $1/\rho$.
We adjust the reservation utilities in order to obtain
\begin{equation*}
0 < \tilde \Phi_{12} < \tilde \Phi_{22}
< \tilde \Phi_{21} = \tilde \Phi_{11}
< \rho \, \tilde \Phi_{12}.
\end{equation*}
To find reservation utilities which achieve this ordering, one can for instance solve the following linear system for $\underline{\tilde u}_1, \underline{\tilde u}_2, \underline{\tilde v}_1, \underline{\tilde v}_2$:
\begin{align*}
\tilde \Phi_{11} - \underline{\tilde u}_1 - \underline{\tilde v}_1
&= 1 + \rho/2 \\
\tilde \Phi_{12} - (1/\rho) \, \underline{\tilde u}_1 - \underline{\tilde v}_2
&= 1 \\
\tilde \Phi_{21} - \underline{\tilde u}_2 - \underline{\tilde v}_1
&= 1 + \rho/2 \\
\tilde \Phi_{22} - \underline{\tilde u}_2 - \underline{\tilde v}_2
&= 1 + \rho/4.
\end{align*}
This system is invertible and therefore always has a solution.

After all these adjustments, we obtain a problem as depicted in figure \ref{fig:nonexchangeability_proof}.
The two following outcomes are stable for this subproblem:
on the one hand, $\mu_{12} = \mu_{21} = 1$, $\tilde u_1 = \rho \, \tilde \Phi_{12}$, $\tilde u_2 = \tilde \Phi_{21}$, and $\tilde v_1 = \tilde v_2 = 0$ (black dots in figure \ref{fig:nonexchangeability_proof});
and on the other hand, $\mu_{11}' = \mu_{22}' = 1$, $\tilde u_1' = \tilde u_2' = 0$, $\tilde v_1' = \tilde \Phi_{21}$, and $\tilde v_2' = \tilde \Phi_{22}$ (white dots in figure \ref{fig:nonexchangeability_proof}).
These solutions are also not exchangeable since, for instance, $\tilde v_2' = \tilde \Phi_{22} > \tilde \Phi_{12}$ is not feasible under the matching $\mu_{12} = 1$.

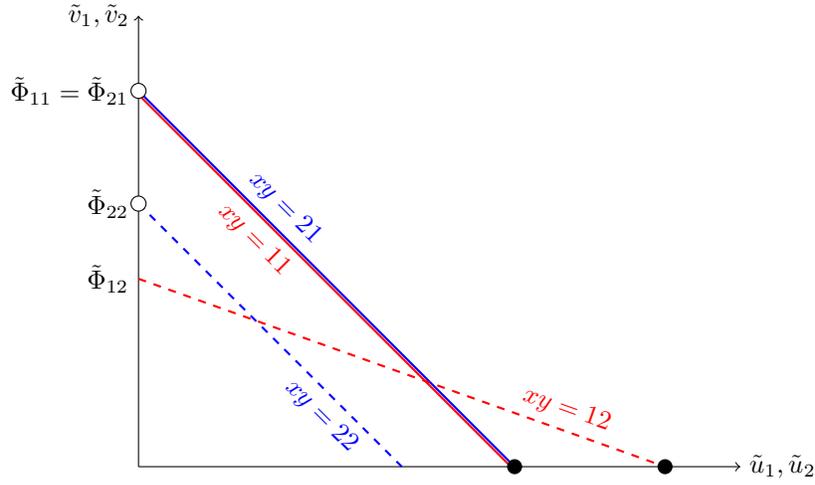
\begin{figure}[h]
\centering
\begin{tikzpicture}
    \draw[->] (0,0) -- (8,0) node[right] {$\tilde u_1, \tilde u_2$};
    \draw[->] (0,0) -- (0,6) node[left] {$\tilde v_1, \tilde v_2$};
    \draw[thick, blue] (0,5) -- (5,0)
        node[pos=0, left, black] {$\tilde \Phi_{11} = \tilde \Phi_{21}$}
        node[pos=0.35, sloped, above] {$xy=21$};
    \draw[thick, blue, dashed] (0,3.5) -- (3.5,0)
        node[pos=0, left, black] {$\tilde \Phi_{22}$}
        node[pos=0.75, sloped, below] {$xy=22$};
    \draw[thick, red] (0,4.95) -- (4.95,0)
        node[pos=0.35, sloped, below] {$xy=11$};
    \draw[thick, red, dashed] (0,2.5) -- (7,0)
        node[pos=0, left, black] {$\tilde \Phi_{12}$}
        node[pos=0.8, sloped, above] {$xy=12$};

    \fill (7,0) circle (.1);
    \fill (5,0) circle (.1);

    \draw[fill=white] (0,3.5) circle (.1);
    \draw[fill=white] (0,5) circle (.1);
\end{tikzpicture}
\caption{A subproblem with non-exchangeable solutions.}
\label{fig:nonexchangeability_proof}
\end{figure}

Thus, if the LTU problem does not have the TU property, then it has a subproblem with two non-exchangeable solutions.
\end{proof}

\end{document}